\documentclass[epj,referee]{svjour}
\usepackage{graphics}
\usepackage{amsmath}
\usepackage{amssymb}
\usepackage{times}
\usepackage{epsf}
\usepackage{graphicx}
\usepackage{dcolumn}
\usepackage{bm}
\usepackage{CJK}
\begin{document}
\title{Engineering the accurate distortion of an object's temperature-distribution signature}
\authorrunning{Chen et al.}
\titlerunning{Engineering the accurate distortion of an object's temperature-distribution signature}
\author{Yixuan Chen, Xiangying Shen, \and Jiping Huang
\thanks{\emph{email:} jphuang@fudan.edu.cn}%
}                     
\institute{Department of Physics, State Key Laboratory of Surface Physics, and Collaborative Innovation Center of Advanced Microstructures, Fudan University, Shanghai 200433, China}
\date{Received: date / Revised version: date}
%
\abstract{
It is up to now a challenge to control the conduction of heat. Here we develop a method to  distort the temperature distribution signature of an object at will. As a result, the object  accurately exhibits the same temperature distribution signature as another object that is predetermined, but actually does not exist in the system. Our finite element simulations confirm the desired  effect for different objects with various geometries and compositions. The underlying mechanism lies in the effects of  thermal metamaterials designed by using this method. Our work is of value for applications in thermal engineering.
} 
\maketitle
\section{Introduction}
\label{intro}
Thermal energy is not only everywhere in nature, but also an outcome of many other types of energy like electrical energy, solar energy, nuclear energy, and mechanical energy. Therefore, it is  particularly important to control heat transfer at will. However, it is up to now a challenge to control the conduction of heat because this conduction obeys the diffusion equation, a partial differential equation describing density dynamics in a material with diffusion~\cite{Mai99}. In 2008, Fan {\it et al.}~\cite{FanAPL08} first adopted the coordinate transformation approach to propose a class of thermal metamaterials with novel thermal properties that cannot be found in nature or chemical compound; their work provides a different way to steer heat conduction. As a result, a lot of thermal metamaterials with novel thermal properties have come to appear, such as cloaks (which are used to let heat flow around an object as if the object does not exist)~\cite{FanAPL08,LiJAP10,GuenneauOE12,NarayanaPRL12,HanSR13,MaNPGAM13,EanHinOoi13,SchittnyPRL13,NarayanaAPL13,DedeAPL13,YangJPD13,HeAPL13,HanPRL14,XuPRL14,HanAM2014,GaoEPL13,MaldovanN13}, concentrators (which are used to concentrate heat into a specific region)~\cite{GuenneauOE12,NarayanaPRL12},  inverters (which are used to apparently let heat flow from the region of low temperature to the region of high temperature)~\cite{FanAPL08,GuenneauOE13}, and rotators (which are used to rotate the flow of heat  as if it comes from a different angle)~\cite{NarayanaPRL12,GuenneauOE13}.

In the area of heat conduction,  a temperature distribution signature can be used to identify an object under temperature gradient. In this sense, if one can distort the temperature distribution signature of Object A into that of Object B by using some methods (note here Object B is predetermined, and actually does not exist in the system), Object A will no longer be found according to the detected  temperature distribution signature. In other words, Object A is mistakenly believed to be Object B, thus yielding a type of thermal illusion (the thermal counterpart of optical illusion~\cite{LaiPRL09}).

In this work, we shall develop the coordinate transformation approach for heat conduction~\cite{FanAPL08,LiJAP10}, and propose a kind of  device made of thermal metamaterials, which causes Object A to accurately possess the same temperature distribution signature as Object B; see Fig.~1.  Our finite element simulations in two dimensions confirm the desired  effect for different objects with various geometries and compositions. This kind of device paves a different way for controlling  heat conduction as expected.

\section{Theory}
\label{sec:1}

To proceed, we plot Fig.~1, which schematically shows the thermal illusion under our consideration. In detail,  Fig.~1(a) displays the temperature distribution signature of a pencil (Object A);   Fig.~1(b) depicts the temperature distribution signature of a key (Object B); Fig.~1(c) is  same as Fig.~1(a), but we add an illusion device. As a result, with the help of the illusion device, the  {\it pencil} outside the illusion device in Fig.~1(c) apparently has the same temperature distribution signature as the {\it key} in Fig.~1(b), thus yielding the thermal illusion. Clearly, the illusion device is the key to obtain this kind of thermal illusion.
But, how to design the illusion device? We offer relevant details in Fig.~2.  Fig.~2 just shows  the illusion device  occupying Regions 2-3 embedded in a background of  thermal conductivity  $\kappa_0$, which can be used to  produce the desired thermal illusion: the medium  of $\kappa_1$ occupying Region 1  apparently produces the same temperature distribution signature as the medium   of $\kappa_4$ occupying Region 4. Here Regions 1-2 both have a shape of trapezoid,  Region 3 contains one rectangle and two triangles, and Region 4 is in the shape of hexagon, which has the same shape, size and location as the total area of Regions 1-3. For achieving this kind of thermal illusion, the medium of $\kappa_2$ (complementary medium) occupying Region\,2 thermally cancels the  medium occupying Region\,1, and the  medium of $\kappa_3$ (restoring medium) occupying Region\,3   thermally takes place of the medium occupying Region\,4.  Then, we are in a position to obtain $\kappa_2$ and $\kappa_3$, in order to obtain the illusion device.
Without loss of generality, we consider  the heat conduction equation for the steady state
\begin{equation}
{\bf \triangledown}\cdot(\kappa{\bf \bigtriangledown} T)=0,\label{eq1}
\end{equation}
where $\kappa$ is the thermal conductivity and $T$ is the temperature.
Since Equation~(\ref{eq1}) remains form invariant under coordinate transformations, we may directly apply the coordinate transformation approach to Equation~(\ref{eq1})~\cite{FanAPL08}. As a result, the new thermal conductivity $\kappa '$ in the transformed coordinates  satisfies the following relation~\cite{FanAPL08,LiJAP10},
\begin{equation}
\kappa '=\frac{J\kappa J^t}{{\rm det}(J)},\label{eq2}
\end{equation}
where $J$ is the Jacobian transformation matrix between the original and distorted coordinates,  $J^t$ is the transposed matrix of $J$, and ${\rm det}(J)$ is the determinant of $J$.
As shown in Fig.\,2(a), Region 2 should be used to thermally cancel Region 1, which can be achieved by folding the geometry of Region 1 into Region\,2. So, according to Equation~(\ref{eq2}), the thermal conductivity of Region\,2,  $\kappa_{2}$, is given by
\begin{equation}
\kappa_{2}=\frac{J_{12}\kappa_{1}J_{12}^{t}}{{\rm det}(J_{12})}.\label{eq3}
\end{equation}
Here, $J_{12}$  is  the Jacobian transformation matrix  which is determined by the coordinates transformation between Region 1 [with coordinates ($x_1$, $y_1$)] and Region 2 [with coordinates ($x_2$, $y_2$)],
\begin{equation}
J_{12}=\left(
         \begin{array}{cc}
           \frac{\partial x_{2}}{\partial x_{1}} & \frac{\partial x_{2}}{\partial y_{1}} \\
           \frac{\partial y_{2}}{\partial x_{1}} &  \frac{\partial y_{2}}{\partial y_{1}} \\
         \end{array}
       \right).\label{eq4}
\end{equation}

On the other hand, since Region\,3 of Fig.\,2(a) is used  to replace the whole area corresponding to  Region\,4 of Fig.\,2(b), according to Equation~(\ref{eq2}), the thermal conductivity of Region\,3, $\kappa_{3}$, is given by
\begin{equation}
\kappa_{3}=\frac{J_{43}\kappa_{4}J_{43}^{t}}{{\rm det}(J_{43})},\label{eq5}
\end{equation}
where $J_{43}$ is  the Jacobian transformation matrix  determined by the coordinates transformation between Region 3 [with coordinates ($x_3$, $y_3$)] and Region 4 [with coordinates ($x_4$, $y_4$)],
\begin{equation}
J_{43}=\left(
         \begin{array}{cc}
           \frac{\partial x_{3}}{\partial x_{4}} & \frac{\partial x_{3}}{\partial y_{4}} \\
           \frac{\partial y_{3}}{\partial x_{4}} & \frac{\partial y_{3}}{\partial y_{4}} \\
         \end{array}
       \right).\label{eq6}
\end{equation}

\section{Results}
\label{sec:2}
In order to show the desired thermal illusion convincingly, we are in a position to perform finite element simulations (based on commercial software COMSOL Multiphysics).
For simulations, we set the following coordinates transformation between Region~1 ($x_1$, $y_1$) and Region~2 ($x_2$, $y_2$),
\begin{eqnarray}
x_2 &=& -\frac{x_1}{2},\label{eq7}\\
y_2 &=& y_1.\label{eq8}
\end{eqnarray}
Then, the substitution of Equations~(\ref{eq7})-(\ref{eq8}) into Equations~(\ref{eq3})-(\ref{eq4}) yields $\kappa_2$.
On the other hand, the coordinates transformation between  Region 4 ($x_4$, $y_4$) and Region\,3 ($x_3$, $y_3$) is set to be
\begin{eqnarray}
 y_3 &=& y_4\, {\rm for\,\,the\,\,whole\,\,area\,\,of\,\,Region\,\,3},\label{eq9}\\
 \frac{1}{4}\ &=& \frac{-2y_3+1-x_3}{-2y_4+1-x_4}\,\, {\rm for\,\,the\,\,upper\,\,obtuse\,\,triangle}, \label{eq10}\\
 \frac{1}{4}\ &=& \frac{2y_3+1-x_3}{2y_4+1-x_4}\,\,  {\rm for\,\,the\,\,lower\,\,obtuse\,\,triangle},\label{eq11}\\
 \frac{1}{4}\ &=& \frac{0.2-x_3}{0.2-x_4}\,\,  {\rm for\,\,the\,\,rectangle\,\,of\,\,Region\,\,3}.\label{eq12}
\end{eqnarray}
So, plugging Equations~(\ref{eq9})-(\ref{eq12}) into Equations~(\ref{eq5})-(\ref{eq6}) gives  $\kappa_3$.
Fig.~3 shows our simulation results; in each panel of Fig.~3, the temperature at the left and right boundary is respectively set to be 400 and 300\,K. In Fig.~3(a), Region 1 within the background of  $\kappa_0=40$\,W/\,(m$\cdot$K) is fully occupied by Object A of $\kappa_A=1$\,W/(m$\cdot$ K) in the shape of trapezoid; In Fig.~3(b), Region 4  within the same background is fully occupied by  Object B of $\kappa_B=400$\,W/(m$\cdot$ K) in the shape of hexagon; Fig.~3(c) is  same as Fig.~3(a), but involves an illusion device that occupies Regions 2-3. Clearly, Fig.~3(b) and Fig.~3(c) show the same temperature distribution signature outside either Region 4 of Fig.~3(b) or Regions 1-3 of Fig.~3(c), thus yielding the desired thermal illusion.

Then, we discuss the case of different shapes and compositions. Fig.~3(d-f) is  same as Fig.~3(a-c), but Object A (or Object B)  becomes  an elliptical (or rectangular ) object of  $\kappa_A=400$\,W/(m$\cdot$ K) (or $\kappa_B=1$\,W/(m$\cdot$ K)), which lies in the background and  partially occupies Region 1 (or Region 4). Clearly,  Fig.~3(f) is the thermal illusion of Fig.~3(e) due to the remote control of the illusion device. Similarly, Fig.~3(i) is the thermal illusion of Fig.~3(h) because of  the remote control of the illusion device as well, where the shape of Object A and Object B has been exchanged.

In Fig.~3(a-i), both Region 1 and Region 4 are fully or partially occupied by Object A and Object B. Fig.~3(j-l) shows a different case, where Object A [Fig.~3(j)] is partially occupied by Region 1 instead and Object B [Fig.~3(k)] is outside Region 4. Also, Fig.~3(l) is still the thermal illusion of Fig.~3(k). In other words,  the illusion device proposed in this work can be used to get thermal illusions not only for the whole area of Object A [Fig.~3(c,f,i)], but also for a part of Object A [Fig.~3(l)].

According to Fig.~3, we may turn to the conclusion that this illusion device is capable of realizing various kinds of thermal illusion, which are independent of the shape and composition of Object A and Object B. But, what is the underlying mechanism? As mentioned in Fig.~2, this mechanism lies in the complementary and restoring effects of thermal metamaterials. In particular,  thermal conductivities of these thermal metamaterials should be not only anisotropic, but also negative, as implied by Equations~(\ref{eq3}) and (\ref{eq5}). Since a negative thermal conductivity of a material means that heat flows from the region of low temperature to the region of high temperature, in reality one must apply an external work to make the physics similar to that of negative conductivities as required by  the second law of thermodynamics. In fact,  an electric refrigerator is just  a sort of ``material'' whose effective thermal conductivity is apparently negative because heat is brought from the interior of the refrigerator (namely, the region of low temperature) to the exterior of the refrigerator (i.e., the region of high temperature) due to the input of electric power. On the same footing, for experimental demonstration of our proposal about thermal illusions, Peltier effects~\cite{BakkerPRL10,KovalevEPL12} may help to get apparently negative values of thermal conductivities. In this direction, we plot Fig.~4 that shows a feasible approach on how to obtain heat flux flowing in a specific region from low temperature to high temperature  by acting external work appropriately as if the thermal conductivity were negative. In other words, the presence of external sources makes the physics similar to that of negative conductivities.


\section{Conclusions}
To sum up, by using the coordinate transformation approach for heat conduction~\cite{FanAPL08,LiJAP10}, we have proposed a kind of thermal illusion device composed of thermal metamaterials.  Our finite element simulations in two dimensions have confirmed the desired  thermal illusions for different objects with various geometries and compositions: With the help of the illusion device, Object A could accurately exhibit the same temperature distribution signature as Object B (a predetermined object) although Object B actually does not exist.
The underlying mechanism originates from the  effects of  thermal metamaterials designed by us.  This work not only proposes a concept of thermal illusion for applications in various fields including military use, but also offers  a  method to control  heat flow at will (e.g., it can even be developed for treating the patterns of local heat flux~\cite{zhu2015}).\\ \\

We acknowledge the financial support by the National Natural Science Foundation of China under Grant No. 11222544, by the Fok Ying Tung Education Foundation under Grant No. 131008, by the Program for New Century Excellent Talents in University (NCET-12-0121), and by the Chinese National Key Basic Research Special Fund under Grant No. 2011CB922004.


%

%
%
%

\clearpage
\newpage

{\bf Figure captions}

Fig.~1. Schematic graph showing the concept of thermal illusion proposed in this work; the bar on the left (or right) with ``High temperature'' (or ``Low temperature'') denotes  the heat source of high (or low) temperature. (a) The blue gradient denotes the temperature distribution signature of a pencil (an object); (b) the orange gradient represents the temperature distribution signature of a key (another object); (c)  same as (a), but we add an illusion device that is indicated by the white area. As a result, the temperature distribution signature outside the illusion device is different from that in (a), but same as that in (b). In other words, with the help of the illusion device, the  {\it pencil} outside the illusion device apparently has the same temperature distribution signature as the {\it key} in (b), thus yielding a thermal illusion.

Fig.~2. Schematic graph showing how to design the illusion device occupying Regions 2 and 3 embedded in a background of  thermal conductivity  $\kappa_0$, for producing the thermal illusion, namely, the medium  of $\kappa_1$ occupying Region 1  apparently produces the same temperature distribution signature as the medium   of $\kappa_4$ occupying Region 4. Here Regions 1-2 both have a shape of trapezoid,  Region 3 contains one rectangle and two triangles, and Region 4 is in the shape of hexagon, which has the same shape, size and location as the total area of Regions 1-3. For achieving this kind of thermal illusion, the medium of $\kappa_2$ (complementary medium) occupying Region\,2 thermally cancels the  medium occupying Region\,1, and the  medium of $\kappa_3$ (restoring medium) occupying Region\,3   thermally takes place of the medium occupying Region\,4.  Note that the origin of coordinates is located at the center of the right boundary of Region 1.

Fig.~3. Simulation  results of temperature distribution signatures in two dimensions. Regions 1-4 of Fig.~2 apply to (a)-(l) herein as well, and dashed lines indicate the position of both Region 1 in (d,f,g,i,j,l)  and  Region 4 in (e,h,k). (a) Region 1 within the background of  $\kappa_0=40$\,W/\,(m$\cdot$K) is fully occupied by Object A of $\kappa_A=1$\,W/(m$\cdot$ K) in the shape of trapezoid, (b) Region 4  within the same background is fully occupied by  Object B of $\kappa_B=400$\,W/(m$\cdot$ K) in the shape of hexagon, and (c) is  same as (a), but involves an illusion device that occupies Regions 2-3. (b) and (c) show the same temperature distribution signature outside either Region 4 of (b) or Regions 1-3 of (c), thus yielding the desired thermal illusion. (d-f) is  same as (a-c), but Object A (or Object B)  becomes  an elliptical (or rectangular ) object of  $\kappa_A=400$\,W/(m$\cdot$ K) (or $\kappa_B=1$\,W/(m$\cdot$ K)), which lies in the background and  partially occupies Region 1 (or Region 4). (g-i) is same as (d-f), but Object A (or Object B) has a shape of rectangle (or ellipse) instead. (j) shows the background with a 0.2\,m-width thermal wall of $\kappa_A=1$\,W/\,(m$\cdot$K), which separates the space into two disconnected parts. In (j), the thermal wall just serves as Object A, which contains Region 1. (k) shows Region~4, which only involves the same material as the background. That is, (k) contains Object B, which is actually the truncated thermal wall and lies outside Region 4.(l) is same as (j), but we add an illusion device to Regions 2-3. Similarly, (k) and (l) show the same temperature distribution signature outside either Region 4 of (k) or Regions 1-3 of (l), thus yielding the thermal illusion as well.  The white area in (c), (f), (i), and (l) denotes the temperature whose value exceeds the bounds of color bars.

Fig.~4. Simulation results of  temperature distribution signatures  in two dimensions. The left and right boundaries of the whole simulation box are set to be 400\,K and 300\,K, respectively. (a) shows an rectangular region (ranged from $x=-0.1\,$m to 0.1\,m) of $\kappa_1=-80$\,W/\,(m$\cdot$K) within the background of $\kappa_0=100$\,W/\,(m$\cdot$K). The geometry of (b) is as same as (a), but we change $\kappa_1$ to be $80$\,W/\,(m$\cdot$K) and set the temperature at $x=-0.1$\,m or  $0.1$\,m to be 327.3\,K or 372.7\,K by acting external work appropriately. Clearly, (b) has the same temperature distribution signature as (a), thus showing a feasible method on how to realize apparently negative thermal conductivities in reality.

\clearpage
\newpage

\begin{figure}
\resizebox{2\columnwidth}{!}{%
  \includegraphics{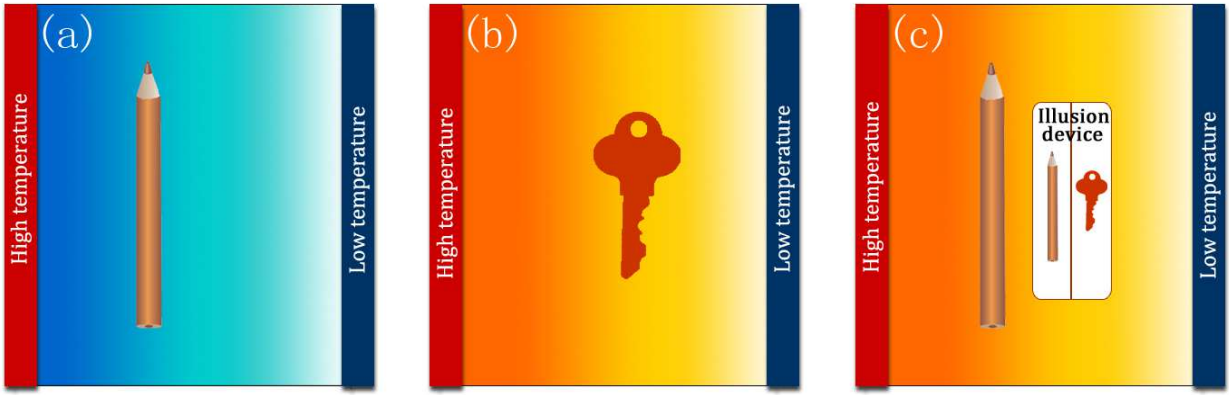}
}
\begin{center}
\caption{/Chen, Shen, Huang}
\label{fig:1}
\end{center}
\end{figure} 

\clearpage
\newpage

\begin{figure}
\resizebox{2\columnwidth}{!}{%
  \includegraphics{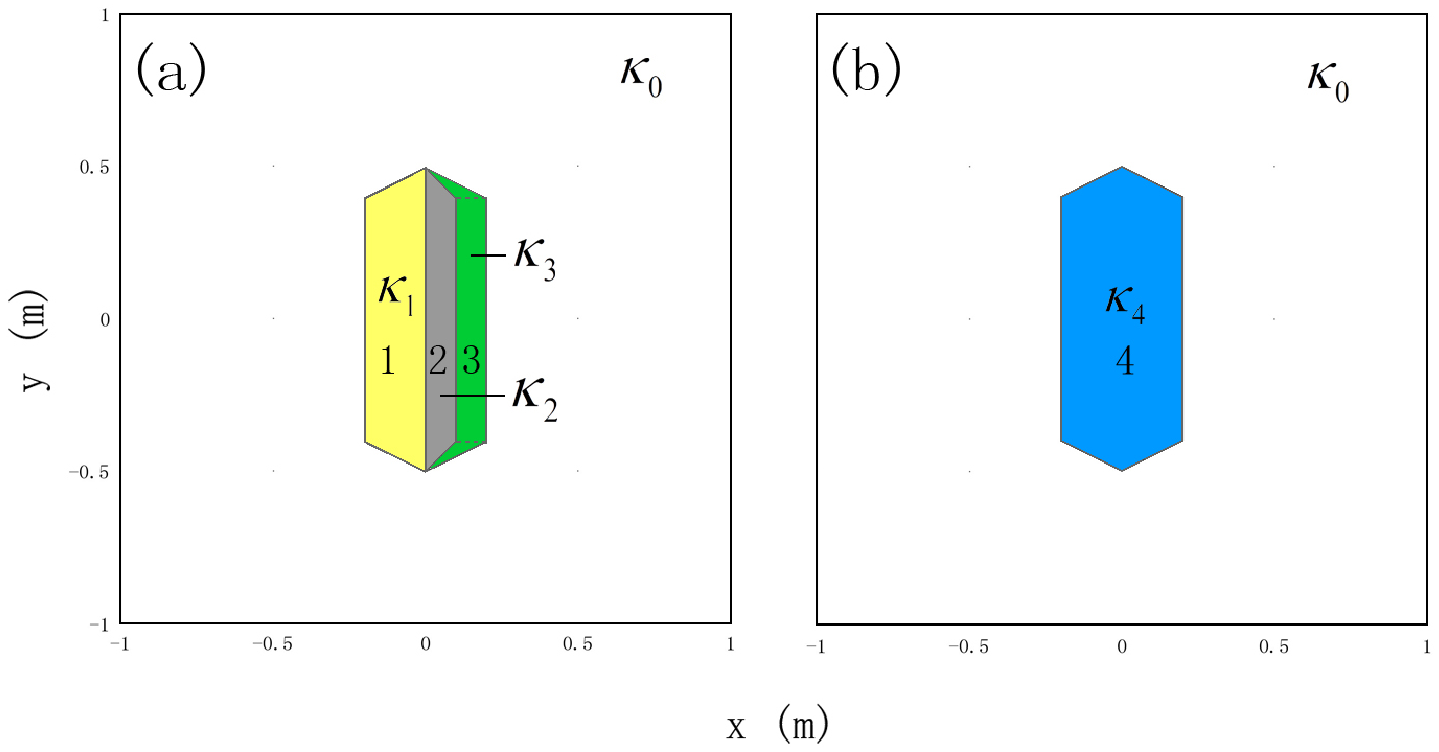}
}
\caption{/Chen, Shen, Huang}
\label{fig:2}       
\end{figure}

\clearpage
\newpage

\begin{figure*}
\begin{center}
\resizebox{1.8\columnwidth}{!}{%
  \includegraphics{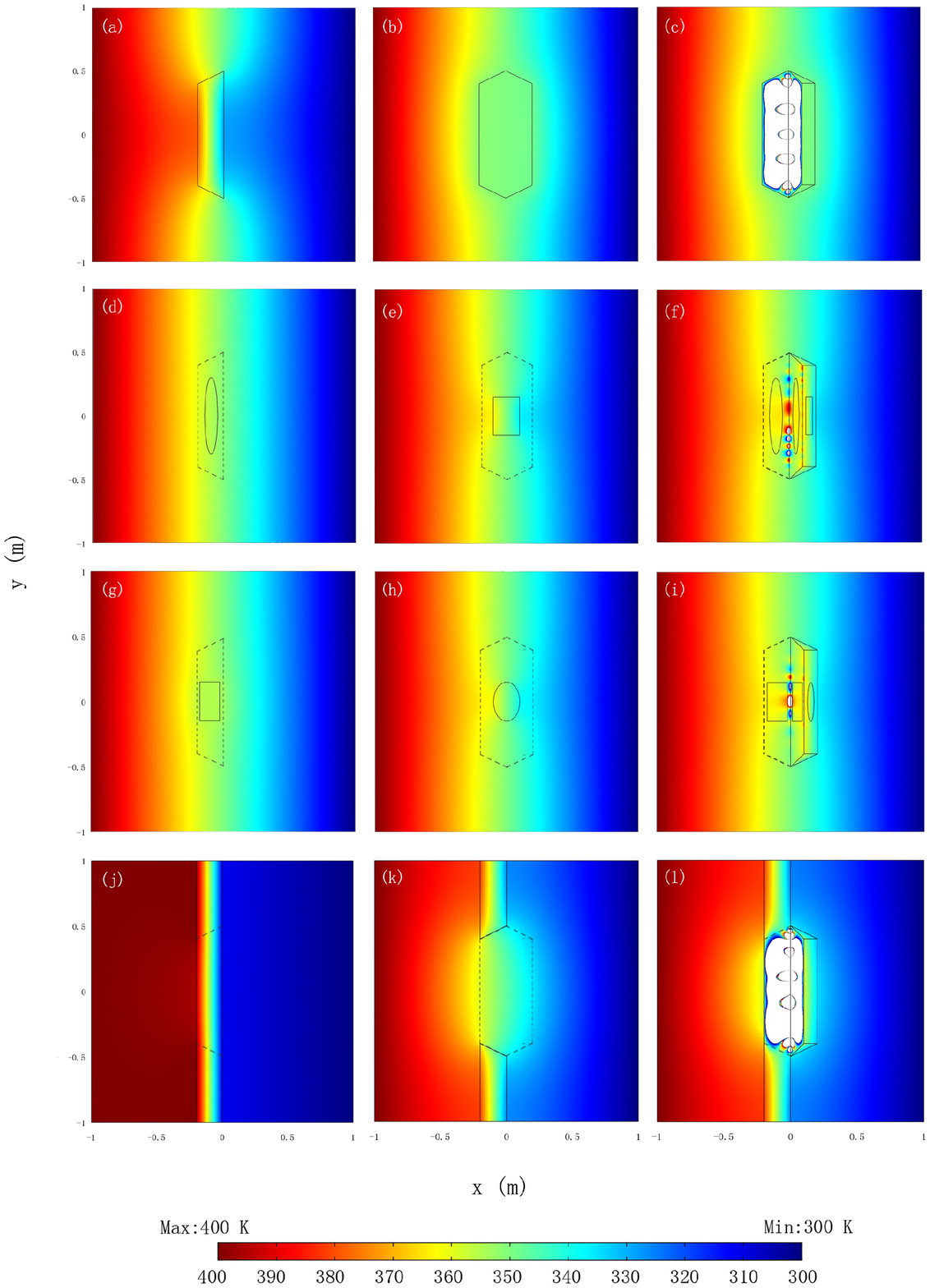}
}
\end{center}      
\caption{/Chen, Shen, Huang}
\label{fig:3}       
\end{figure*}
\clearpage
\newpage

\begin{figure*}
\begin{center}
\resizebox{1.8\columnwidth}{!}{%
  \includegraphics{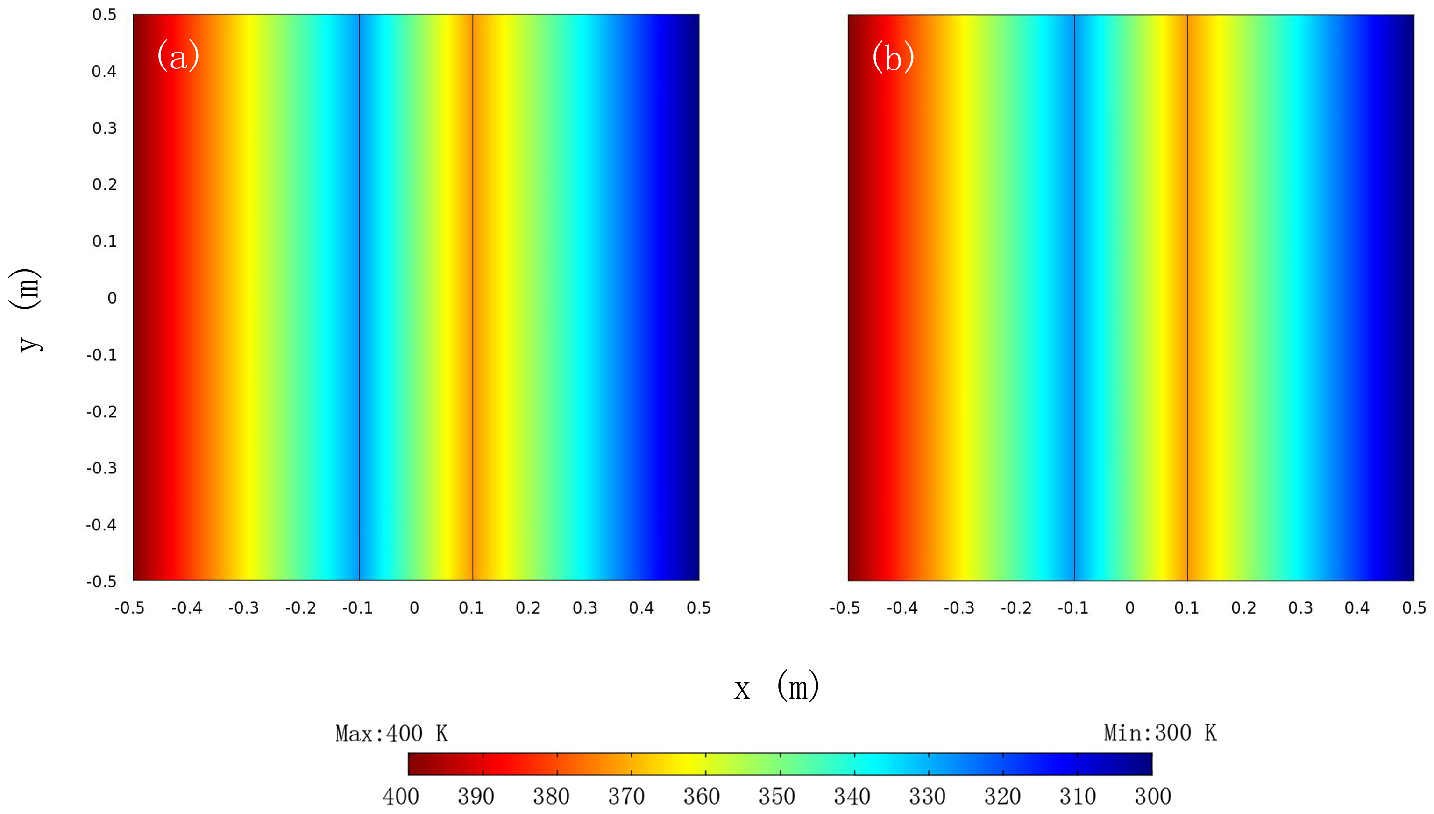}
}
\end{center}      
\caption{/Chen, Shen, Huang}
\label{fig:4}       
\end{figure*}

\clearpage

\end{document}